\documentclass[letter,bibyear]{aa} 

\usepackage{epsfig}
\usepackage{amsmath}
\usepackage{subfigure}
\usepackage{natbib}
\usepackage{multirow}
\usepackage{color}
\usepackage{longtable}
\usepackage{graphicx}
\usepackage{epstopdf}
\usepackage{booktabs}
\usepackage{txfonts}

\newcommand{\cpl}{Chem. Phys. Lett.}

\newcommand{\jms}{J.~Mol.~Spectrosc.}   
  
\newcommand{\jmst}{J.~Mol.~Struct.}

\newcommand{\kms}{km s$^{-1}$}

\begin{document}

\title{Discovery of thiofulminic acid with the QUIJOTE$^1$ line survey: 
A study of the isomers of HNCS and HNCO in TMC-1\thanks{Based
on observations with the Yebes 40m radio telescope (projects 19A003, 20A014, 20D023, 21A011, 21D005, 
22A007, 22B029, and 23A024) and the IRAM 30m radiotelescope. The 40m radiotelescope at Yebes Observatory is operated by the Spanish Geographic Institute (IGN, Ministerio de Transportes y Movilidad Sostenible).
IRAM is supported by INSU/CNRS (France), MPG (Germany), and IGN (Spain).}}

\author{
J.~Cernicharo\inst{1},
M.~Ag\'undez\inst{1},
C.~Cabezas\inst{1},
B.~Tercero\inst{2,3},
R.~Fuentetaja\inst{1},
N.~Marcelino\inst{2,3}, and
P.~de~Vicente\inst{3}
}

\institute{Dept. de Astrof\'isica Molecular, Instituto de F\'isica Fundamental (IFF-CSIC),
C/ Serrano 121, 28006 Madrid, Spain. \newline \email jose.cernicharo@csic.es
\and Observatorio Astron\'omico Nacional (OAN, IGN), C/ Alfonso XII, 3, 28014, Madrid, Spain.
\and Centro de Desarrollos Tecnol\'ogicos, Observatorio de Yebes (IGN), 19141 Yebes, Guadalajara, Spain.
}

\date{Received: 9 December 2023; accepted: 9 January 2024}

\abstract{We present the first detection of HCNS (thiofulminic acid)  in space 
with the QUIJOTE$^1$ line survey in the direction of TMC-1.
We performed a complete study of the isomers of CHNS and CHNO, including NCO and NCS. The derived column densities
for HCNS, HNCS, and HSCN are (9.0$\pm$0.5)$\times$10$^9$, (3.2$\pm$0.1)$\times$10$^{11}$, and (8.3$\pm$0.4)$\times$10$^{11}$ cm$^{-2}$, respectively. The HNCS/HSCN abundance ratio
is 0.38. 
The abundance ratios HNCO/HNCS, HCNO/HCNS, HOCN/HSCN, and NCO/NCS are
34$\pm$4, 8.3$\pm$0.7, 0.18$\pm$0.03, and 0.78$\pm$0.07, respectively. These ratios cannot be correctly 
reproduced by our gas-phase chemical models, which suggests that formation paths for these species are missing,
and/or that the adopted dissociative recombination rates for their protonated precursors have to be revised.
The isotopologues H$^{15}$NCO, DNCO, 
HN$^{13}$CO, DCNO, H$^{34}$SCN,
and DSCN have also been detected with the ultrasensitive QUIJOTE line survey. 
}

\titlerunning{Thiofulminic acid}
\authorrunning{Cernicharo et al.}

\maketitle

\section{Introduction}
Among the new species discovered in recent years with the QUIJOTE\footnote{\textbf{Q}-band \textbf{U}ltrasensitive \textbf{I}nspection \textbf{J}ourney
to the \textbf{O}bscure \textbf{T}MC-1 \textbf{E}nvironment} line survey of TMC-1 
\citep[][and references therein]{Cernicharo2021a,Cernicharo2023a,Cernicharo2023b},
it is worth noting that 11 of them are S-bearing species: NCS, HCCS, H$_2$CCS, H$_2$CCCS, C$_4$S,
HCSCN, HCSCCH, HC$_4$S, HC$_3$S$^+$, HCCS$^+$, and HSO \citep{Cernicharo2021b,Cernicharo2021c,Cernicharo2021d,Cabezas2022a,
Fuentetaja2022,Marcelino2023}. The chemical models for S-bearing molecules in interstellar clouds poorly reproduce the
observed abundances and the results are strongly dependent on the adopted 
depletion of sulfur \citep{Vidal2017}. Moreover,
many reactions involving S$^+$ with neutrals, as well as radicals
with S and CS, are missing in the chemical networks and further study is required in order to achieve better 
agreement between chemical models and observations of sulfur-bearing species \citep{Petrie1996,Bulut2021}.
In addition to these new species, other S-bearing molecules such as CS, SO, NS, NS$^+$, HCS$^+$, HSC, HCS, 
CCS, C$_3$S, C$_5$S, H$_2$CS, HSCN, and HNCS are present in TMC-1 \citep[See Table 1 in][]{Cernicharo2021b}.
Despite the high depletion of sulphur in molecular clouds, most of these species are the counterpart 
of the same molecular structure but replacing sulphur with oxygen. Only C$_4$O, HCCO$^+$, and HC$_4$O have not yet been 
detected in TMC-1, or indeed in any other source. HNCS \citep{Frerking1979} and HSCN \citep{Halfen2009} are isomers of the CHNS family, and
obvious counterparts of the isomers of CHNO. For the latter family, HNCO \citep{Snyder1972}, 
HCNO \citep{Marcelino2009}, and HOCN \citep{Brunken2009a,Marcelino2010} have
been detected in space in a broad sample of cold and warm molecular clouds. 

The study of isomers of the same molecule provides
clues as to the formation mechanism of these species, as often they are produced from a
common precursor. 
The recent detection of five cyano derivatives
of propene, which have energies ranging from 0 to 19.6 kJ/mol (0-2357 K) but similar abundances, indicates
that kinetics, and not energetics, is dominating the production of these different isomers \citep{Cernicharo2022a}.
The most relevant isomers in space are HCN and HNC, which are assumed to be produced from the dissociative electronic
recombination ($DR$) of HCNH$^+$. In cold dark clouds, these isomers have an abundance ratio of $\sim$1 in spite
of the much higher energy of HNC  of 55 kJ/mol 
\citep[][and references therein]{Baulch2005,Mendes2012}. 
Their chemistry in interstellar clouds
was analysed in detail by \citet{Loison2014}.

Concerning the CHNS and CHNO isomers, it is assumed that the first are mainly formed from $DR$ of H$_2$NCS$^+$ and HNCSH$^+$
\citep{Adande2010,Gronowski2014}, 
and the second from $DR$ of H$_2$NCO$^+$ and HNCOH$^+$ \citep{Marcelino2010,Quan2010}.
In the present work, we present the first detection of HCNS (thiofulminic acid)  in space (TMC-1) and a coherent and
precise study of the abundances of the isomers of CHNS and CHNO. The isotopologues
H$^{34}$SCN, DSCN, H$^{15}$NCO, HN$^{13}$CO, DNCO, and DCNO are also detected in TMC-1. The derived abundances for the 
CHNS and CHNO isomers, together with those of NCS and NCO,
are compared with the results of state-of-the-art chemical models for S-bearing species.

\begin{figure}[h] 
\centering
\includegraphics[width=0.61\textwidth]{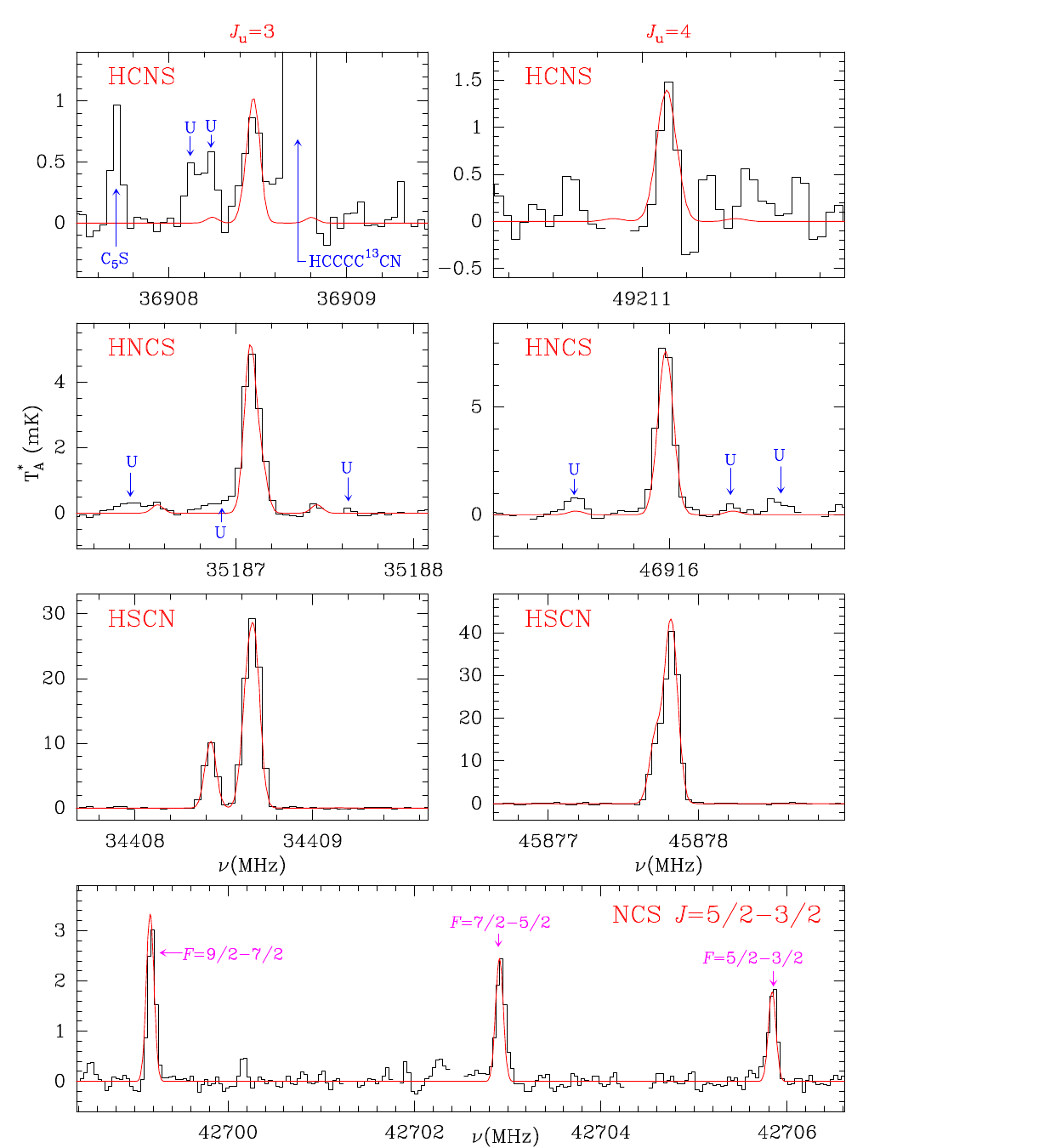}
caption{Observed $Ju=3$ and $4$ ($K_a=0$) transitions of the isomers of HCNS. The bottom panel
shows the hyperfine components of the $J=5/2-3/2$ transition of NCS in its $^2\Pi_{3/2}$ ground
state.
The abscissa corresponds to the rest frequency. The ordinate is the antenna temperature corrected for 
atmospheric and telescope losses, in mK.
Blanked channels correspond to negative features produced when folding the frequency-switched data.
The red line shows the computed synthetic spectra for these lines (see Sect. \ref{sec:results}). 
The derived line parameters
are given in Table \ref{line_parameters_hcns}.
} 
\label{fig:thios}
\end{figure}

\section{Observations}
The observational data used in this work are part of QUIJOTE \citep{Cernicharo2021a}, 
a spectral line survey of TMC-1 in the Q-band carried out with the Yebes 40m telescope at 
the position $\alpha_{J2000}=4^{\rm h} 41^{\rm  m} 41.9^{\rm s}$ and $\delta_{J2000}=
+25^\circ 41' 27.0''$, corresponding to the cyanopolyyne peak (CP) in TMC-1. 
A detailed description of the system 
is given by \citet{Tercero2021}, and details on the QUIJOTE line survey are provided by \citet{Cernicharo2021a,Cernicharo2023a,Cernicharo2023b}.
The data analysis procedure is described by \citet{Cernicharo2022b}.
The total observing time on source is 1202 hours and the 
measured sensitivity varies between 0.07 mK at 32 GHz and 0.2 mK at 49.5 GHz.  

The main beam efficiency 
can be given across the Q-Band by
$B_{\rm eff}$=0.797 exp[$-$($\nu$(GHz)/71.1)$^2$]. The
forward telescope efficiency is 0.95 and
the beam size at half power intensity is 54.4$''$ and 36.4$''$ 
at 32.4 and 48.4 GHz, respectively.

In addition to the  Q-band data, the isomers of HNCO and HNCS were observed in the millimetre domain with 
the IRAM 30m telescope and consist of a 3mm line survey 
that covers the 71.6-116 GHz domain 
\citep{Cernicharo2012,Agundez2022,Cabezas2022a,Cernicharo2023b}. 
The final 3mm line survey has a sensitivity of 2-10 mK. However, at some selected 
frequencies, the sensitivity is as low as 0.6 mK. 
The IRAM 30m beam 
varies between 34$''$ and 21$''$ at 72 GHz and 117 GHz, respectively, while the beam 
efficiency takes values of 0.83 and 0.78 at the same frequencies (B$_{eff}$= 0.871 
exp[$-$($\nu$(GHz)/359)$^2$]). The forward efficiency at 3mm is 0.95.

The absolute intensity calibration uncertainty is 10$\%$. However, the relative
calibration between lines within the QUIJOTE survey is certainly better because all
them are observed simultaneously and have the same calibration uncertainties.
The data were analysed with the GILDAS package\footnote{\texttt{http://www.iram.fr/IRAMFR/GILDAS}}.

\section{Results}\label{sec:results}

Line identification in this work was performed using the MADEX code \citep{Cernicharo2012b}, and the CDMS 
and JPL catalogues \citep{Muller2005,Pickett1998}.
The references for the laboratory data used by MADEX in its spectroscopic predictions are described below.
The intensity scale 
used in this study is the antenna temperature ($T_A^*$). Consequently, the telescope parameters and source 
properties were
used when modelling the emission of the different species in order to produce synthetic spectra in this
temperature scale. The permanent dipolar moments of the molecular species studied in this work are
given in Table \ref{columndensities}. In this work, we assumed a velocity for the source relative to the
local standard of rest of 5.83 \kms\, \citep{Cernicharo2020}. The source is assumed to be circular
with a uniform brightness temperature and a radius of 40$''$ \citep{Fosse2001}.
The procedure to derive line parameters is described in Appendix \ref{app:lineparameters}.
The observed line intensities were modelled using a local thermodynamical equilibrium
hypothesis (LTE), or a large
velocity gradient approach (LVG). In the latter case, MADEX uses the formalism described by \citet{Goldreich1974}.

\begin{table}
\centering
\caption{Derived column densities for the isomers of HNCO and HNCS}
\label{columndensities}
\centering
\begin{tabular}{{lcccc}}
\hline
Species             & T$_{rot}$  & $\mu_a$    & $N^A$                        & Notes\\
                    &  (K)       &  (D)      & (cm$^{-2}$)                &      \\
\hline                           
HCNS                & 7.0        &  3.85 & (9.0$\pm$0.5)$\times$10$^{09}$& a\\
HNCS                & 6.0        &  1.64 & (3.2$\pm$0.1)$\times$10$^{11}$& b\\
HSCN                & 5.2        &  3.33 & (8.3$\pm$0.4)$\times$10$^{11}$& c\\
HSNC                & 5.2        &  3.13 & $\le$4$\times$10$^{09}$       & a\\
NCS                 & 5.5        &  2.45 & (9.5$\pm$0.2)$\times$10$^{11}$& d\\
HNCO                & 8.0        &  1.58 & (1.1$\pm$0.1)$\times$10$^{13}$& d\\
HCNO                & 5.5        &  3.10 & (7.8$\pm$0.2)$\times$10$^{10}$& e\\
HOCN                & 5.5        &  3.70 & (1.5$\pm$0.2)$\times$10$^{11}$& f\\
HONC                & 5.5        &  3.13 & $\le$5$\times$10$^{09}$       & g\\
NCO                 & 8.0        &  0.64 & (7.4$\pm$0.5)$\times$10$^{11}$& h\\
H$_2$NCO$^+$        & 5.5        &  4.13 & (1.8$\pm$0.4)$\times$10$^{10}$& i\\
\hline
\hline
H$^{34}$SCN         & 5.2        &  3.33 & (2.1$\pm$0.3)$\times$10$^{10}$& j\\
DSCN                & 5.2        &  3.33 & (2.1$\pm$0.3)$\times$10$^{10}$& j\\
H$^{15}$NCO         & 8.0        &  1.58 & (3.3$\pm$0.3)$\times$10$^{10}$& j\\
HN$^{13}$CO         & 8.0        &  1.58 & (2.1$\pm$0.1)$\times$10$^{11}$& j\\
DNCO                & 8.0        &  1.58 & (2.5$\pm$0.2)$\times$10$^{11}$& j\\
DCNO                & 5.5        &  3.10 & (3.1$\pm$0.4)$\times$10$^{09}$& j\\
\hline
\end{tabular}
\tablefoot{
\tablefoottext{A}{Column densities for the isomers of HCNS, together with
the most abundant S-bearing species, were reported by \citet{Cernicharo2021b}. The present
estimations are $\sim$20\% larger due to the recalibration of the
QUIJOTE data in 2023 \citep{Cernicharo2023a}.}
\tablefoottext{a}{$\mu_a$ from \citet{McGuire2016}.}
\tablefoottext{b}{$\mu_a$ from \citet{Szalanski1978}.}
\tablefoottext{c}{$\mu_a$ from \citet{Woon2009}.}
\tablefoottext{d}{$\mu_a$ from CDMS \citep{Muller2005}.}
\tablefoottext{e}{$\mu_a$ from \citet{Takashi1989}.}
\tablefoottext{f}{$\mu_a$ from \citet{Brunken2009a}.}
\tablefoottext{g}{$\mu_a$ from \citet{Mladenovic2009}.}
\tablefoottext{h}{$\mu_a$ from \citet{Saito1970}.}
\tablefoottext{i}{$\mu_a$ from \citet{Gupta2013}. 
}
\tablefoottext{j}{Assumed identical to that of the main isotopologue.}

}
\end{table}

\subsection{The HCNS, HSCN, and HNCS isomers}\label{sect:thios}
The HCNS, HNCS, and HSCN isomers were observed with high spectral resolution in the
laboratory \citep{Brunken2009b,McGuire2016}. The frequencies measured in these studies were fitted
using the adapted Hamiltonian for each species and the resulting rotational constants were implemented into the
MADEX code to predict transition frequencies and to model the observed emission line profiles.

HNCS and HSCN are well-known molecular components of
interstellar clouds \citep{Frerking1979,Halfen2009,Adande2010,Vastel2018}. They harbour strong lines
in our line survey. However, the isomer HCNS has not been detected
yet in space. It is the thio-form of HCNO, an isomer of HNCO and a molecule detected in several interstellar
clouds \citep{Marcelino2009}. Thanks to the sensitivity of QUIJOTE, 
we report in this work the first detection of its two lines in the
Q-band ($J=3-2$ and $J=4-3$)  in space, namely towards TMC-1. The lines are detected with a signal-to-noise ratio (S/N)
of $\sim$12 and
match the predicted frequencies within 3 kHz (see Fig. \ref{fig:thios}). The
derived line parameters are given in Table \ref{line_parameters_hcns}. The sensitivity of our
line survey at 3mm with the IRAM 30m telescope is not as high as that of QUIJOTE. Nevertheless, the
$J=6-5$ line of HCNS at 73815.849 MHz is detected at a 3$\sigma$ level in the data (see Fig. \ref{fig:HCNS_6}). We also observed this line with the Yebes 40m telescope using the W-band receiver described by \citet{Tercero2021}. Near the frequency
of the HCNS $J=6-5$ transition, there is a line of CCC$^{13}$CH that is detected with good S/N
with both telescopes. It has an intensity of 10 mK and has been used to scale the data of the Yebes 40m and the IRAM 30m telescopes. The two sets of data were merged, weighting them as 1/$\sigma^2$. The
resulting data allow us to detect the $J=6-5$ line of HCNS
at 4$\sigma$. The results are shown in Fig. \ref{fig:HCNS_6}. We conclude that HCNS is convincingly
detected in TMC-1. A fit to the laboratory and astronomical frequencies
for this species provides the following new molecular constants: $B$=6151.44367$\pm$0.00027 MHz, $D$=1.712$\pm$0.016 kHz, 
and $eQq$=-0.7227$\pm$0.0018 MHz.

From the observed intensities of these three lines of HCNS, and using a line profile
fit method \citep{Cernicharo2021d},
we obtain T$_{rot}$=7.0$\pm$0.5\,K and N(HCNS)=(9.0$\pm$0.5)$\times$10$^9$
cm$^{-2}$. The synthetic spectra are shown in Fig. \ref{fig:thios} and \ref{fig:HCNS_6}, and are in
excellent agreement with the observations. No collisional rates are available for HCNS, but with the aim to
estimate the excitation conditions that could be expected for this molecule in the LVG approximation,
we adopted the collisional
rates of OCS with H$_2$ calculated by \citet{Green1978}. OCS has a rotational constant similar
to that of HCNS, and therefore we consider that these rates provide a better approximation
to those of HCNS than the rates of HCNO \citep{Naindouba2015}, which has a rotational constant twice that of HCNS.
For the typical volume densities of TMC-1 of 1$\times$10$^4$ cm$^{-3}$ \citep[][Fuentetaja et al., in prep.]{Agundez2023},
we obtain that the excitation temperatures of the observed lines could vary between 8.2\,K for the 
$J=3-2$ line down to 6.2\,K for the $J=6-5$ transition, in good agreement with the derived 
averaged rotational temperature of 7\,K.

The $J_u=3,4$ with $K_a=0$ transitions of HNCS and HSCN appear in our QUIJOTE 
data as features exhibiting all their hyperfine components with a very high S/N. These transitions
are shown in Fig. \ref{fig:thios}. 
The corresponding $K_a=1$ lines of HNCS have upper energy levels of 66 ($J_u=3$) and 68\,K ($J_u=4$)
and are not detected in our data. However, the corresponding energies for HSCN are only at 16.9 and 19.2\,K,
respectively. We detected these latter; they are shown in Fig. \ref{fig:thios_hscn}. Moreover, the 
$J_u=7, 8$ transitions with $K_a$=0 of both isomers were detected with
the IRAM 30m telescope (see Fig. \ref{fig:thios_hscn}). The derived line parameters for all these
lines are given in Table \ref{line_parameters_hcns}. No collisional rates are available for either of
the two isomers. We used all observed lines using the line profile fitting procedure \citep{Cernicharo2021d}
to derive their rotational temperature and column
densities. For HNCS, we obtain T$_{rot}$=6.0$\pm$0.5\,K and N(HNCS)=(3.2$\pm$0.1)$\times$10$^{11}$ cm$^{-2}$.
For HSCN, we derived T$_{rot}$=5.2$\pm$0.3\,K and N(HSCN, $K_a$=0) = (6.9$\pm$0.2)$\times$10$^{11}$ cm$^{-2}$,
and N(HSCN, $K_a$=1) = (1.4$\pm$0.3)$\times$10$^{11}$ cm$^{-2}$. The interladder temperature between
$K_a$=0 and 1 lines could be thermalised by collisions at the kinetic temperature of the cloud. For an energy separation between these ladders of 14.5\,K, we derive a kinetic temperature for TMC-1 of 9.1$\pm$1.5\,K, which is in excellent agreement with the result obtained by \citet{Agundez2023}.

The isomer HSNC has been characterised in the laboratory \citep{McGuire2016} and we searched for its
lines in our survey. However, we only obtained an upper limit of 4$\times$10$^9$ cm$^{-2}$  for its column density. Of particular interest for the chemistry of the CHNS isomers, it is worth deriving the column density of NCS as 
it could participate in their formation. NCS was discovered in TMC-1 with previous QUIJOTE line data
\citep{Cernicharo2021b}. The
new data are shown in Fig. \ref{fig:thios}. Adopting a rotational temperature of 5.5\,K, the derived
column density is (9.5$\pm$0.2)$\times$10$^{11}$ cm$^{-2}$. The NCS to HNCS abundance ratio is 3.0$\pm$0.16.

Among the isomers of the CHNS family, the most abundant is HSCN. Its lines are strong enough to allow a
search for its isotopologues. We observed two rotational transitions exhibiting hyperfine structure
for H$^{34}$SCN and DSCN (see Fig. \ref{fig:iso_hscn}). Adopting the same rotational temperature as for
the main isotopologue, we derive the column densities given in Table \ref{columndensities}. The $^{32}$S/$^{34}$S
abundance ratio is 33$\pm$5, which agrees with previous determinations using other molecular species (see e.g.
\citealt{Cernicharo2021b} and
Fuentetaja et al. in prep., and references therein). The H/D abundance ratio is also 33$\pm$5 and represents
a considerable single deuterium enhancement in HSCN, in line with that observed in other molecules in TMC-1
\citep[][ and references therein]{Cabezas2021,Cabezas2022b}.

\subsection{The HNCO, HCNO, and HOCN isomers}
HNCO was one of the first molecules detected in the early years of millimetre radio astronomy by \citet{Snyder1972}. Its isomer
HCNO (fulminic acid) was detected by \citet{Marcelino2009} towards B1, L1544, and L183, but not in TMC-1. Another isomer, HOCN, was detected towards SgrB2, B1-b, L1544, L1527, L183, and IRAS16293-2422
\citep{Brunken2009a,Brunken2010,Marcelino2010}. Therefore, these isomers have 
been detected in space in a broad sample of cold and warm molecular clouds. The level energies, 
transition frequencies, and intensities for HNCO and its isotopologues, as well as for HOCN, are from the JPL and
CDMS catalogues, respectively. HCNO is a quasi-linear molecule and its rotational frequencies were obtained from a fit to available laboratory data \citep{Winnewisser1971}. The three isomers and the isotopologues $^{15}$N, $^{13}$C, and D of HNCO have been detected with QUIJOTE in TMC-1. The data are shown in Fig. \ref{fig:HNCO} and the derived line
parameters are given in Table \ref{line_parameters_hnco}. We note that for HN$^{13}$CO and DNCO, our measured
frequencies have a significant lower uncertainty than those in the entries of the JPL catalogue.

The HNCO lines are the strongest among those of all its isomers. We also detected the 
$4_{04}-3_{03}$ and $5_{05}-4_{04}$ lines at 3mm with the IRAM 30m radio telescope. These
lines are shown in Fig. \ref{fig:o-3mm}. Collisional rates for HNCO with He (T$_K\ge$30\,K;
\citealt{Green1986}) and H$_2$ (T$_K\ge$7\,K; \citealt{Sahnoun2018}) are available. 
For a volume density of 1-2$\times$10$^{4}$, our LVG calculations using the $p$-H$_2$
collisional rates provide excitation temperatures for the $2_{0,2}-1_{0,1}$ line of HNCO of 8-9\,K,
decreasing to $\sim$5\,K for the $5_{05}-4_{04}$ transition. We therefore adopted a rotational
temperature of 8\,K for HNCO and its isotopologues in the Q-band, and of 4.8\,K in the 3mm domain. The derived column densities are given in
Table \ref{columndensities} and the computed synthetic spectra are shown in red in Figures \ref{fig:HNCO}
and \ref{fig:o-3mm}. The HNCO/HN$^{13}$CO, HNCO/DNCO, and HNCO/H$^{15}$NCO abundance ratios are
52$\pm$8, 44$\pm$8, and 333$\pm$60. 

\begin{table*}
\centering
\caption{Derived abundance ratios for the isomers of HNCO and HNCS in TMC-1}
\label{tab:ratios}
\centering
\begin{tabular}{{|lc|cc|c|}}
\hline
S-species      & N(X)/N(HNCS) & O-species & N(X)/N(HNCO) & N(S)/N(O) \\
\hline                           
HCNS           &   0.028$\pm$0.002   & HCNO        & 0.0071$\pm$0.0008 & 0.12$\pm$0.01 \\
HNCS           &   1.0               & HNCO        & 1.0               & 0.029$\pm$0.004 \\
HSCN           &   2.60$\pm$0.016    & HOCN        & 0.014$\pm$0.003   & 5.52$\pm$0.91 \\
HSNC           &  $\le$0.013         & HONC        & $\le$0.0005       &  \\
NCS            &   3.00$\pm$0.16     & NCO         & 0.067$\pm$0.011   & 1.28$\pm$0.11 \\
NS$^a$         &   5.3$\pm$0.7       & NO$^b$      & 25$\pm$6              & 0.007$\pm$0.002 \\
               &                     & H$_2$NCO$^+$& 0.0016$\pm$0.0005 &                     \\
\hline
\end{tabular}
\tablefoot{
\tablefoottext{a}{Column density of (1.7$\pm$0.2)$\times$10$^{12}$ cm$^{-2}$ \citep{Cernicharo2018}.}
\tablefoottext{b}{Column density of (2.7$\pm$0.5)$\times$10$^{14}$ cm$^{-2}$ \citep{Gerin1993}.}
}
\end{table*}

The isomers HCNO and HOCN have dipolar moments of 3.1 and 3.7\,D, respectively 
(see Table \ref{columndensities}). For HCNO we used the LVG approximation using 
the collisional rates of HCNO with He calculated by \citet{Naindouba2015}. For a density
of 1-2$\times$10$^4$ cm$^{-3}$, we obtain rotational temperatures in the range of 4.9-5.5\,K. 
HCNO is detected for the first time in TMC-1. We also observed its $J=4-3$ transition at 3mm (see Fig. 
\ref{fig:thios_hscn}). The estimated column density for this isomer of HNCO is (7.8$\pm$0.2)$\times$10$^{10}$ cm$^{-2}$.
We also observed the $J=2-1$ of DCNO in the Q-band (see Fig. 
\ref{fig:o-3mm}). The derived HCNO/DCNO abundance ratio
is 25$\pm$4. For HOCN, we observed the
$2_{0,2}-1_{0,1}$ with QUIJOTE and its $4_{04}-3_{03}$ and $5_{05}-4_{04}$ lines at 3mm 
(see Fig. \ref{fig:o-3mm}). 
All the observed lines are well reproduced by the models 
with a rotational temperature similar to that of HCNO ($T_{rot}$=5.5\,K) as shown by the computed synthetic 
spectra shown in Figures \ref{fig:HNCO} and \ref{fig:o-3mm}.
The derived column densities for HCNO and HOCN are given in Table \ref{columndensities}.
The fundamental transition of HOCN at 20.9 GHz
was tentatively observed in TMC-1 by \citet{Brunken2009b}. 

We detected the four strongest hyperfine components of the $J=7/2-5/2$ transition of NCO 
in its $^2\Pi_{3/2}$ ground state at 3mm. This
species was recently detected 
towards L483 by \citet{Marcelino2018} with an NCO/HNCO abundance ratio of $\sim$0.2. Due to the modest
dipole moment of this species \citep[$\mu$=0.64\,D; ][]{Saito1970}, we
modelled the TMC-1 data adopting T$_{rot}$(NCO)=8\,K. The derived column density is (7.4$\pm$0.5)$\times$10$^{11}$ cm$^{-2}$ and the NCO/HNCO abundance ratio is 0.067, which is three times lower than
towards L483. However, in TMC-1 the NCS/HNCS ratio is 45 times larger than the NCO/HNCO one. We also searched for the isomer HONC \citep{Mladenovic2009}, but only obtained upper limits (see Table \ref{columndensities}). 
Finally, three lines of H$_2$NCO$^+$ were observed in the Q-band. These
are shown in Fig. \ref{fig:H2NCO+} and their line parameters are given in Table \ref{line_parameters_hnco}. Adopting
a rotational temperature of 5.5\,K and a $o$/$p$ abundance ratio of 3:1, the
derived total column density for this species is (1.8$\pm$0.4)$\times$10$^{10}$ cm$^{-2}$. 

\begin{figure}
\centering
\includegraphics[width=\columnwidth]{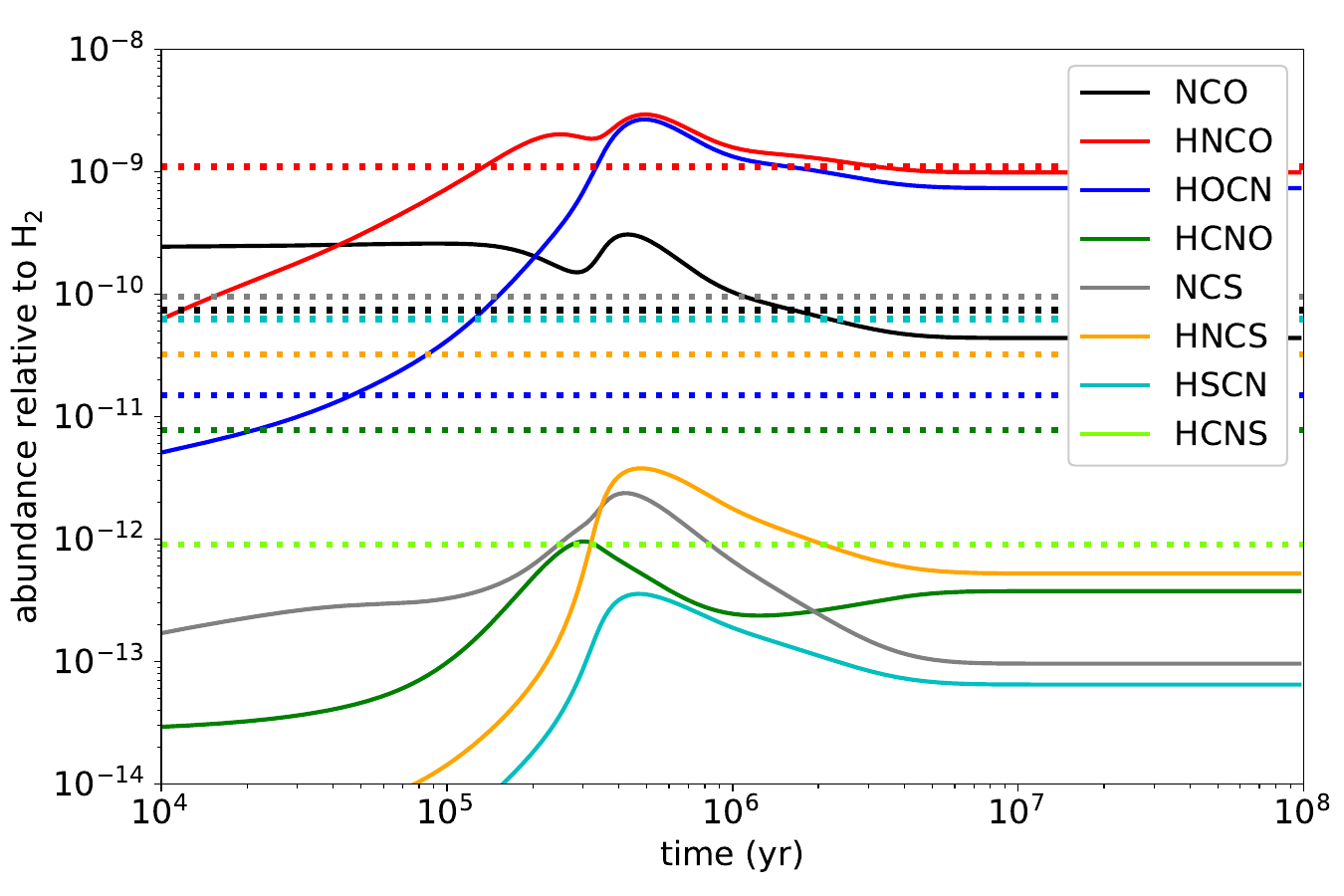}
\caption{Abundances calculated with the chemical model for molecules of the families of HNCO and HNCS. The peak calculated abundance of HCNS is around 10$^{-15}$ relative to H$_2$ and therefore lies outside the abundance range shown. Horizontal dashed lines correspond to the abundances observed in TMC-1.}
\label{fig:abun}
\end{figure}

\section{Discussion}\label{sec:discussion}
The most energetically stable isomer of the CHNS family is HNCS, followed by HSCN, HCNS, and HSNC at 
3171, 17317, and 18120\,K above, respectively \citep{Wierzejewska2003,McGuire2016}. The energy ordering in the oxygen family is similar, with HNCO being the most stable isomer, and HOCN, HCNO, and HONC lying at 
12330, 34480, and 42235\,K above, respectively \citep{Schuurman2003}. In spite of the similarities in the energetics, the CHNO and CHNS families show a very different behaviour in TMC-1. While HNCO is far more abundant than HOCN, the abundance of HNCS is only half of that of HSCN (see Table \ref{tab:ratios}).

The chemistry of HNCS and its isomers has been discussed by \cite{Adande2010}, \cite{Gronowski2014}, and \cite{Vidal2017}. Here, we revisit the subject by running a gas-phase chemical model of a cold dense cloud (see details in \citealt{Cernicharo2021b} and \citealt{Fuentetaja2022}). The chemical scheme of synthesis of HNCS and HSCN can be summarized as
\begin{equation}
\rm XH^+ + NCS \rightarrow HNCS^+ + X,
\end{equation}
\begin{equation}
\rm HNCS^+ + H_2 \rightarrow H_2NCS^+/HNCSH^+ + H, \label{reac2}
\end{equation}
\begin{equation}
\rm H_2NCS^+ + e^- \rightarrow HNCS + H,
\end{equation}
\begin{equation}
\rm HNCSH^+ + e^- \rightarrow HNCS/HSCN + H, \label{reac4}
\end{equation}
where the scheme begins with the protonation of NCS by abundant proton donors XH$^+$, such as HCO$^+$, H$_3$O$^+$, and H$_3^+$. This scheme, replacing sulfur by oxygen, is the same one that describes the formation of the oxygen analogues HNCO and HOCN \citep{Marcelino2010,Quan2010,Lattanzi2012}. The chemical schema start with the radicals NCO and NCS, which are mostly formed through neutral--neutral reactions. According to the chemical model, the main reactions forming NCO are N + HCO and CN + O$_2$, while NCS is mainly formed through the analogous reactions N + HCS and CN + SO.

The abundances calculated with the chemical model are shown in Fig.~\ref{fig:abun}. We focus on times around a few 10$^5$ yr, when most molecules reach their peak abundances. The abundances calculated for NCO and HNCO are in line with the observed values. In the case of HOCN, the calculated abundance is close to that of HNCO, while observations indicate that HOCN is much less abundant than HNCO. It is likely that the dissociative recombination of HNCOH$^+$ favours HNCO over HOCN (the model assumes equal branching ratios) or that the reaction between HNCO$^+$ and H$_2$ favours H$_2$NCO$^+$ over HNCOH$^+$ (equal yields are assumed at present). Investigation into these two reactions would allow us to understand the high observed HNCO/HOCN ratio. Grain surface chemistry may also be responsible for this. Indeed, the gas-grain model of \cite{Quan2010} predicted HNCO to be somewhat more abundant than HCNO. The detection of the precursors H$_2$NCO$^+$ and HNCOH$^+$ would also be very useful to shed light on this. While the H$_2$NCO$^+$ cation has been detected in interstellar clouds \citep{Gupta2013,Marcelino2018}, its isomer HNCOH$^+$ is still awaiting detection in space (laboratory frequencies are available from \citealt{Mladenovic2009}). Only the H$_2$NCO$^+$ cation is detected in our TMC-1 data.

In the case of the sulfur analogues NCS, HNCS, and HSCN, the model severely underestimates their abundances. It is likely that the model lacks efficient routes to NCS, hampering the subsequent formation of HNCS and HSCN. A similar situation was encountered in previous chemical models, with abundances relative to H$_2$ $\lesssim$ 10$^{-12}$ for HNCS
and HSCN \citep{Adande2010,Gronowski2014,Vidal2017}. Apart from the general problem of abundance underestimation, the fact that HSCN is observed to be more abundant than HNCS, in contrast to the oxygen case, indicates that the branching ratios of reactions\,(\ref{reac2}) and (\ref{reac4}) must be markedly different to those of the analogous oxygen reactions. Identification of the precursor cations H$_2$NCS$^+$ and HNCSH$^+$ would also help to constrain the chemistry of HNCS and HSCN, although their rotational spectrum has not yet been characterised in the laboratory.

The third observed species of the family of HNCO isomers, namely HCNO, has a different skeleton to the other two isomers and is therefore formed in a different way. According to the chemical model, the neutral--neutral reaction CH$_2$ + NO \citep{Marcelino2009} is the main source of HCNO. The analogous molecule with sulfur, HCNS (the one reported here), is severely underestimated by the chemical model. The analogous reaction CH$_2$ + NS cannot account for HCNS because in TMC-1, NS is 160 times less abundant than NO, while HCNS is just 9 times less abundant than HCNO
(see Table \ref{tab:ratios}).
Among the potential missing routes to HCNS, 
we could consider the neutral--neutral reactions HCN + SH, CN + H$_2$S, 
N + H$_2$CS, NH + HCS, and NH$_2$ + CS. To our knowledge, these reactions have not been studied either theoretically or experimentally, and therefore we do not know whether or not they are fast at low temperatures and produce HCNS as a product. Among these reactions, the most efficient by far is N + H$_2$CS, which would provide an HCNS abundance of the order of the observed one assuming that it has a rate coefficient of the order of 1$\times$10$^{-10}$ cm$^3$ s$^{-1}$ at low temperature. 

We conclude that further experimental and theoretical work is needed to understand the chemistry of the HNCS and HNCO isomers. 
In particular, the enhancement of the abundance of 
HSCN relative to HNCS and HOCN, as well as the behaviour of the NCS/HNCS and NCO/HNCO abundance ratios, are not 
well reproduced by the present chemical models.

\begin{acknowledgements}
We thank Ministerio de Ciencia e Innovaci\'on of Spain (MICIU) for funding support through projects
PID2019-106110GB-I00, 
and PID2019-106235GB-I00. We also thank ERC for funding
through grant ERC-2013-Syg-610256-NANOCOSMOS.

\end{acknowledgements}

\normalsize
\begin{appendix}
\section{Line parameters}\label{app:lineparameters}
Line parameters for all observed transitions with the Yebes 40m and IRAM 30m radio telescopes
were derived by fitting a Gaussian line profile to them
using the GILDAS package. A
velocity range of $\pm$20\,\kms\, around each feature was considered for the fit after a polynomial 
baseline was removed. Negative features produced in the folding of the frequency switching data were blanked
before baseline removal. 

\begin{figure}[h] 
\centering
\includegraphics[width=0.39\textwidth]{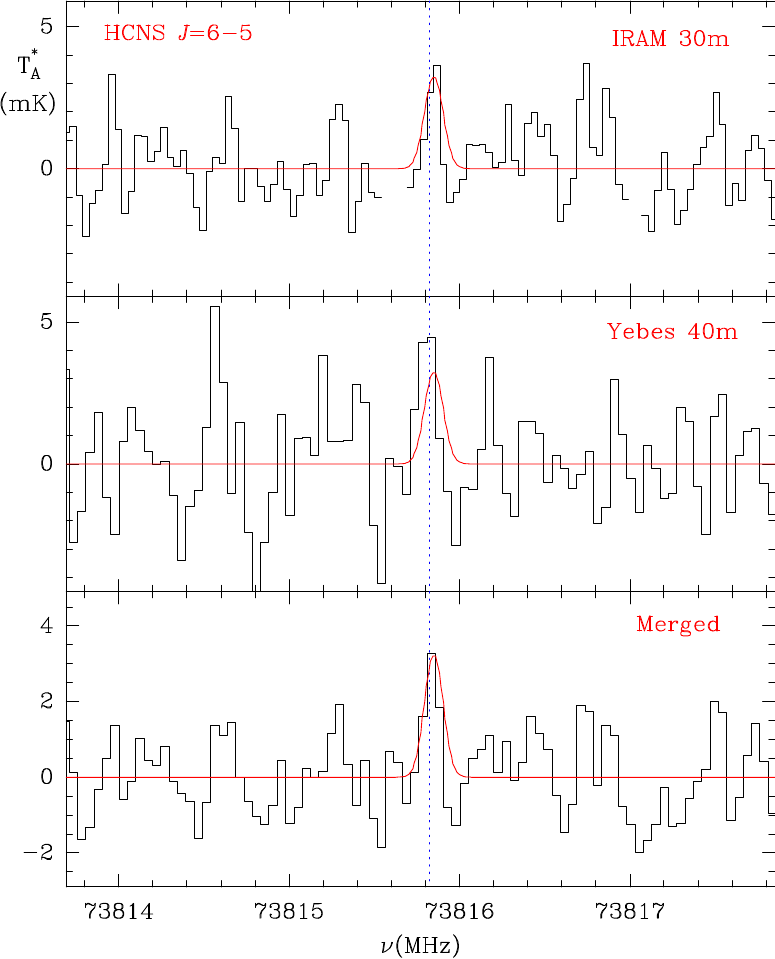}
\caption{Observed $J=6-5$ line of HCNS with the IRAM 30m (top panel) 
and Yebes 40m (middle panel) radio telescopes. The data of the Yebes
telescope have been scaled to those of IRAM using the line
$N=8-7$ $J=15/2-13/2$ of CCC$^{13}$CH at 73819.18 MHz that appears
in both sets of data with a S/N of $\sim$10 in the IRAM data
and of $\sim$5 in the Yebes data. The merged spectrum was obtained
by averaging both sets of data weighted as 1/$\sigma^2$ and has a sensitivity of 1\,mK.
The derived line parameters for this transition of HCNS are given in Table \ref{line_parameters_hcns}.
The red line corresponds to the computed synthetic spectra (see Sect. \ref{sec:results}).
The vertical blue line corresponds to the measured frequency for a v$_{LSR}$ of 5.83\,km\,s$^{-1}$.
} 
\label{fig:HCNS_6}
\end{figure}

\begin{figure}[h] 
\centering
\includegraphics[width=0.59\textwidth]{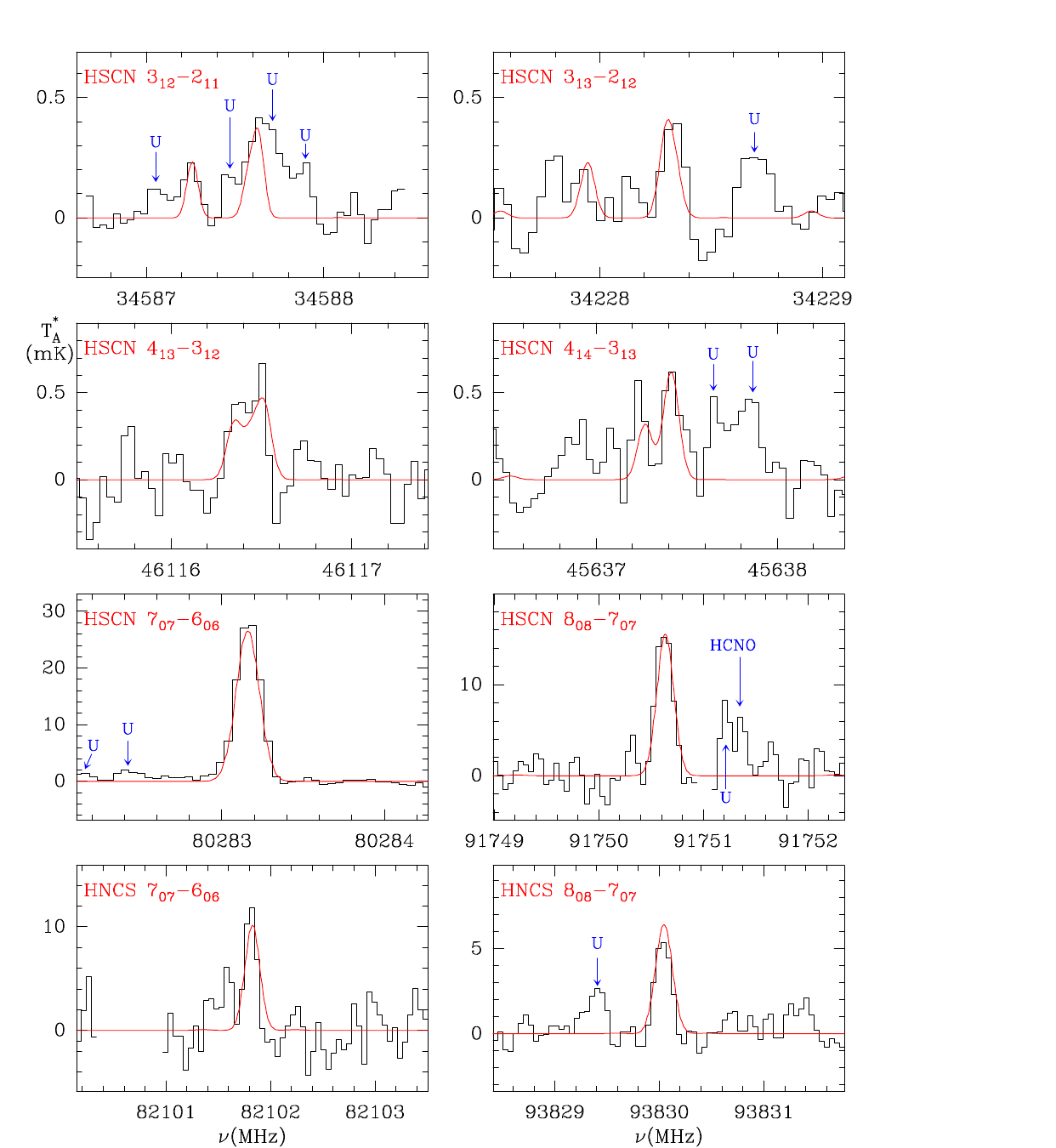}
\caption{Observed lines of HSCN and HNCS.
The two upper rows show the observed lines of HSCN with $K_a$=1.
The two lower rows show the $7_{07}-6_{06}$
and $8_{08}-7_{07}$ lines of HSCN (third row) and HNCS (fourth row)
observed at 3mm with the IRAM 30m radio telescope.
The red line shows the computed synthetic spectra for these lines (see Sect. \ref{sec:results}). 
The derived line parameters
are given in Table \ref{line_parameters_hcns}.
} 
\label{fig:thios_hscn}
\end{figure}

The derived line parameters for all observed lines of the CHNS and CHNO families are
given in Tables \ref{line_parameters_hcns} and \ref{line_parameters_hnco}, respectively. The differences
between the measured frequencies and the predicted ones are always of a few kHz. However, for some
isotopologues, several hyperfine components are blended in the laboratory data 
(measurements with uncertainties $\ge$ 30 kHz), while in our observations
the lines are well resolved and have an uncertainty of 10 kHz.

\begin{figure}[h] 
\centering
\includegraphics[width=0.61\textwidth]{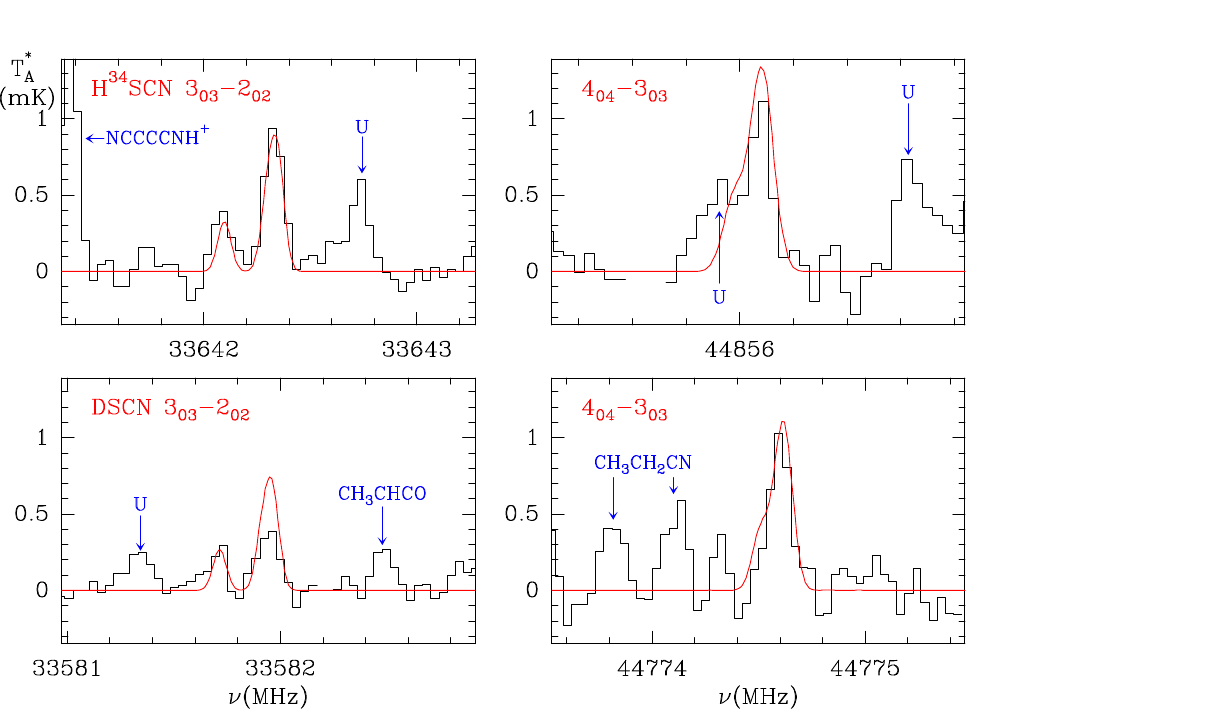}
\caption{Observed $3_{03}-2_{02}$ and $4_{04}-3_{03}$ transitions of the isotopologues H$^{34}$SCN (top panels)
and DSCN (bottom panels) of HSCN.
The red line shows the computed synthetic spectra for these lines (see Sect. \ref{sec:results}). 
The derived line parameters are given in Table \ref{line_parameters_hcns}.
} 
\label{fig:iso_hscn}
\end{figure}

\begin{figure}[h] 
\centering
\includegraphics[width=0.5\textwidth]{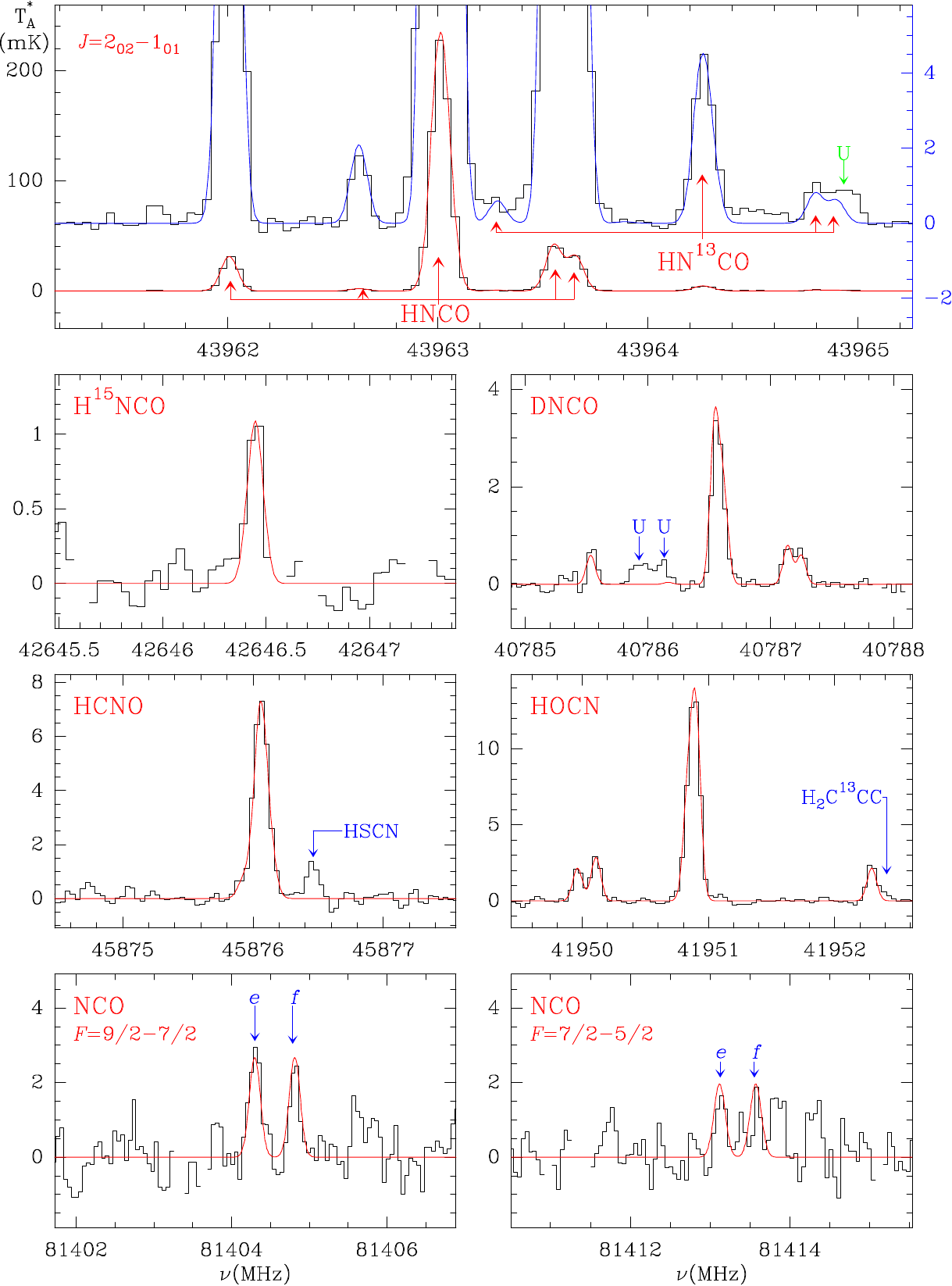}
\caption{Same as Fig. \ref{fig:thios} but for the 
$J=2_{02}-1_{01}$ transition of HNCO (isocyanic acid) and its isotopologues HN$^{13}$CO, H$^{15}$NCO, and DCNO
(upper and second raw panels). The upper panel also shows a zoom to the data to exhibit the hyperfine
lines of HN$^{13}$CO. The scale of the zoom is indicated in blue on the right-Y axis.
The third row shows the $J=2-1$ line of HCNO (the $J=4-3$ is shown in Fig.
\ref{fig:thios_hscn}) 
and the $J=2_{02}-1_{01}$ line of HOCN (cyanic acid).
Finally, the bottom panels show two hyperfine components of the $J$=7/2-5/2 line
of NCO in its $^2\Pi_{3/2}$ ground state. The lambda-type doubling components,
$e$ and $f$, are indicated.
The derived line parameters are given in Table \ref{line_parameters_hnco}.
The hyperfine structure was considered in the modelling of the lines (shown by
the red lines).
} 
\label{fig:HNCO}
\end{figure}

\begin{figure}[h] 
\centering
\includegraphics[width=0.48\textwidth]{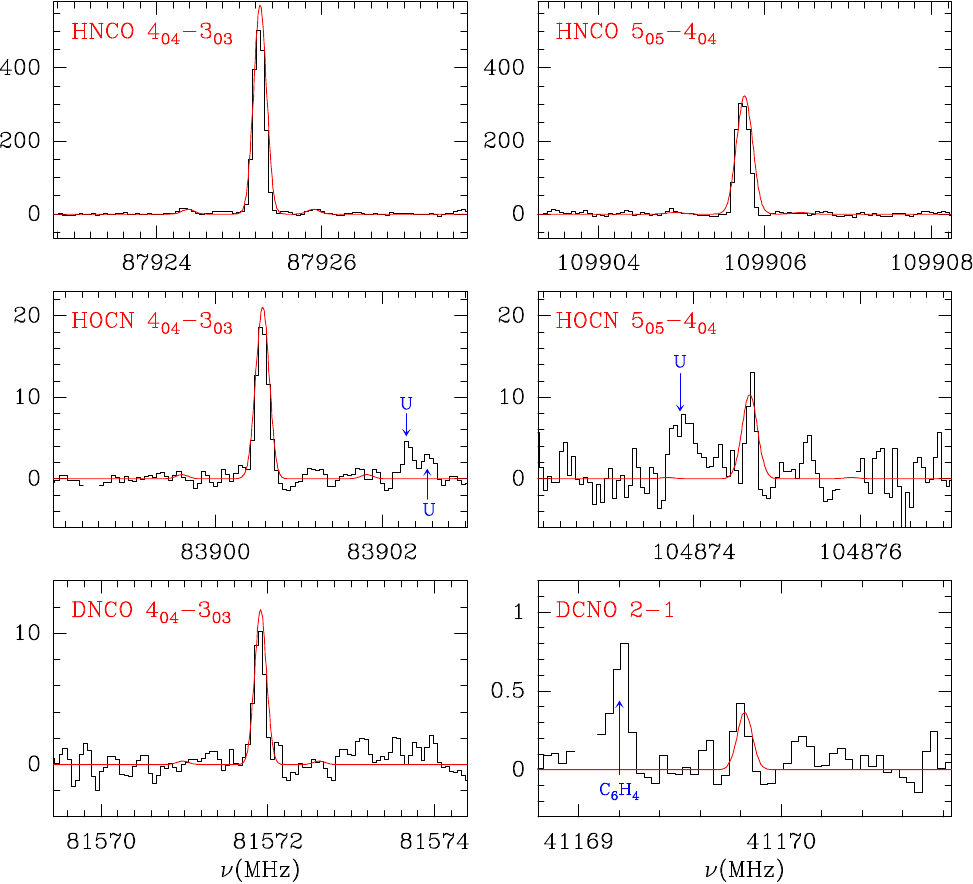}
\caption{Observed lines of HNCO and HOCN at 3mm with the IRAM 30m radio telescope.
The red line shows the synthetic spectra computed 
with the models described in Sect. \ref{sec:results}. The column densities
are the same as those derived from the lines observed in the Q-band. The bottom panels
show the DNCO $4_{04}-3_{03}$ and DCNO $J=2-1$ lines.
The derived line parameters are given in Table \ref{line_parameters_hnco}.
} 
\label{fig:o-3mm}
\end{figure}

\begin{figure}[h] 
\centering
\includegraphics[width=0.49\textwidth]{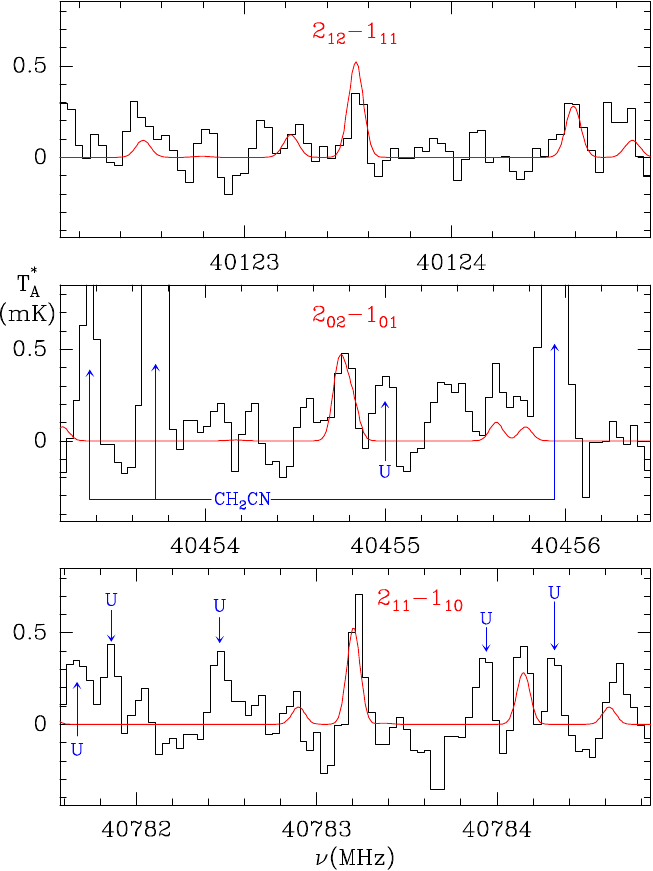}
\caption{Observed $J_u=2$ lines of H$_2$NCO$^+$ towards TMC-1.
The red line shows the computed synthetic spectra for these lines (see Sect. \ref{sec:results}). 
The derived line parameters are given in Table \ref{line_parameters_hnco}.
} 
\label{fig:H2NCO+}
\end{figure}

The lines of HCNS, HNCS, HSCN, and NCS are shown
in Figures \ref{fig:thios} and \ref{fig:thios_hscn}. 
The observations of the $J=6-5$ line of HCNS with the IRAM 30m and the Yebes 40m
radio telescopes, together with the resulting merged data, are shown in Fig. \ref{fig:HCNS_6}.
Finally, the lines of the isotopologues H$^{34}$SCN and
DSCN are shown in Fig. \ref{fig:iso_hscn}.
The lines of HNCO, HCNO, HOCN, and NCO are shown
in Fig. \ref{fig:HNCO} and those of H$_2$NCO$^+$ in Fig. \ref{fig:H2NCO+}

\onecolumn

\begin{table*}[h]
\centering
\caption{Observed line parameters for the isomers and isotopologues of HNCS}
\label{line_parameters_hcns}
\begin{tabular}{lcccrccrc}
\hline
Species    & $J_u-J_l$  &$F_u-F_l$$^a$& $\nu_{rest}$~$^b$ & $\int T_A^* dv$~$^c$ & v$_{LSR}$       & $\Delta v$~$^d$ & $T_A^*$~$^e$& Notes\\
                &                      & & (MHz)              & (mK\,km\,s$^{-1}$)  & (km\,s$^{-1}$)  & (km\,s$^{-1}$)  & (mK) & \\
\hline
HCNS       & 3-2           &         & 36908.475$\pm$0.010& 0.95$\pm$0.06& 5.83 &  1.04$\pm$0.09& 0.85$\pm$0.07& \\
           & 4-3           &         & 49211.115$\pm$0.010& 0.77$\pm$0.11& 5.83 &  0.47$\pm$0.08& 1.56$\pm$0.14& \\
           & 6-5           &         & 73815.835$\pm$0.020& 1.34$\pm$0.39& 5.83 &  0.38$\pm$0.10& 3.40$\pm$0.90& \\
           &               &         &                    &              &      &               &              & \\                 
HNCS       &$3_{03}-2_{02}$&  2-2    & 35186.557$\pm$0.010& 0.19$\pm$0.06& 5.83 &  0.63$\pm$0.10& 0.29$\pm$0.07& \\
           &               & Main    & 35187.089$\pm$0.010& 5.05$\pm$0.06& 5.83 &  0.99$\pm$0.09& 4.82$\pm$0.07& \\
           &               &  3-3    & 35187.445$\pm$0.010& 0.16$\pm$0.04& 5.83 &  0.46$\pm$0.11& 0.33$\pm$0.07& \\
                   &$4_{04}-3_{03}$&  3-3    & 46915.466$\pm$0.010& 0.67$\pm$0.10& 5.83 &  0.72$\pm$0.13& 0.87$\pm$0.14&A\\
           &               & Main    & 46915.976$\pm$0.010& 5.98$\pm$0.11& 5.83 &  0.68$\pm$0.05& 8.19$\pm$0.14& \\
           &               &  4-4    & 46916.349$\pm$0.010& 0.27$\pm$0.10& 5.83 &  0.51$\pm$0.23& 0.50$\pm$0.14&A\\
                   &$7_{07}-6_{06}$& Main    & 82101.810$\pm$0.010& 6.04$\pm$0.09& 5.83 &  0.45$\pm$0.07&12.68$\pm$2.10& \\
                   &$8_{08}-7_{07}$& Main    & 93830.036$\pm$0.010& 3.29$\pm$0.16& 5.83 &  0.53$\pm$0.04& 5.83$\pm$0.60& \\
           &               &         &                    &              &      &               &              & \\
HSCN       &$3_{13}-2_{12}$& 3-2     & 34227.919$\pm$0.020& 0.14$\pm$0.06& 5.83 &  0.62$\pm$0.12& 0.21$\pm$0.09& \\
           &$3_{13}-2_{12}$& Main    & 34228.329$\pm$0.010& 0.39$\pm$0.06& 5.83 &  0.84$\pm$0.14& 0.44$\pm$0.09& \\
           &$3_{03}-2_{02}$& 3-3     & 34407.322$\pm$0.010& 1.62$\pm$0.09& 5.83 &  0.77$\pm$0.04& 1.98$\pm$0.11& \\
           &$3_{03}-2_{02}$& 2-1     & 34408.425$\pm$0.010& 7.74$\pm$0.08& 5.83 &  0.72$\pm$0.01&10.26$\pm$0.11& \\
           &$3_{03}-2_{02}$& Main    & 34408.660$\pm$0.010&28.58$\pm$0.09& 5.83 &  0.90$\pm$0.01&29.82$\pm$0.11& \\
           &$3_{03}-2_{02}$& 2-2     & 34410.448$\pm$0.010& 1.38$\pm$0.09& 5.83 &  0.67$\pm$0.04& 1.93$\pm$0.11& \\
           &$3_{12}-2_{11}$& 3-2     & 34587.252$\pm$0.010& 0.27$\pm$0.05& 5.83 &  1.02$\pm$0.21& 0.24$\pm$0.06& \\
           &$3_{12}-2_{11}$& Main    & 34587.606$\pm$0.010& 0.35$\pm$0.05& 5.83 &  1.53$\pm$0.32& 0.21$\pm$0.06&B\\
           &$4_{14}-3_{13}$& 4-3     & 45637.236$\pm$0.010& 0.28$\pm$0.07& 5.83 &  0.45$\pm$0.14& 0.57$\pm$0.14& \\
           &$4_{14}-3_{13}$& Main    & 45637.423$\pm$0.010& 0.51$\pm$0.09& 5.83 &  0.81$\pm$0.18& 0.59$\pm$0.14& \\
           &$4_{04}-3_{03}$& 4-4     & 45876.458$\pm$0.010& 0.80$\pm$0.07& 5.83 &  0.61$\pm$0.05& 1.24$\pm$0.14& \\
           &$4_{04}-3_{03}$& 3-2     & 45877.725$\pm$0.010&10.52$\pm$0.34& 5.83 &  0.68$\pm$0.02&14.57$\pm$0.14& \\
           &$4_{04}-3_{03}$& Main    & 45877.828$\pm$0.010&27.29$\pm$0.34& 5.83 &  0.65$\pm$0.01&39.75$\pm$0.14& \\
           &$4_{04}-3_{03}$& 3-3     & 45878.621$\pm$0.010& 0.39$\pm$0.08& 5.83 &  0.88$\pm$0.08& 0.41$\pm$0.14& \\
           &$4_{13}-3_{12}$& 4-3     & 46116.344$\pm$0.020& 0.18$\pm$0.04& 5.83 &  0.60$\pm$0.00& 0.30$\pm$0.13&C\\
           &$4_{13}-3_{12}$& 3-2     & 46116.396$\pm$0.020& 0.28$\pm$0.07& 5.83 &  0.60$\pm$0.00& 0.31$\pm$0.13&C\\
           &$4_{13}-3_{12}$& 5-4     & 46116.492$\pm$0.020& 0.36$\pm$0.07& 5.83 &  0.60$\pm$0.00& 0.60$\pm$0.13&C\\
           &$7_{07}-6_{06}$& Main    & 80283.160$\pm$0.010&19.98$\pm$0.44& 5.83 &  0.61$\pm$0.02&28.89$\pm$0.64& \\
           &$8_{08}-7_{07}$& Main    & 80283.160$\pm$0.010&10.54$\pm$0.77& 5.83 &  0.60$\pm$0.05&16.63$\pm$1.66& \\
           &               &         &                    &              &      &               &              & \\
DSCN       &$3_{13}-2_{12}$& Main    & 33258.268$\pm$0.010& 0.24$\pm$0.07& 5.83 &  0.72$\pm$0.21& 0.32$\pm$0.09& \\
           &$3_{03}-2_{02}$& Main    & 33581.957$\pm$0.010& 0.33$\pm$0.04& 5.83 &  0.87$\pm$0.11& 0.35$\pm$0.06&D\\
           &$3_{12}-2_{11}$& Main    & 33905.771$\pm$0.003&              &      &               & $\le$0.3     & \\
           &$4_{14}-3_{13}$& Main    & 44343.744$\pm$0.010& 0.36$\pm$0.12& 5.83 &  0.66$\pm$0.26& 0.52$\pm$0.10& \\
           &$4_{04}-3_{03}$& Main    & 44774.598$\pm$0.010& 0.84$\pm$0.13& 5.83 &  0.78$\pm$0.14& 1.01$\pm$0.13& \\
           &$4_{13}-3_{12}$& Main    & 45207.143$\pm$0.010& 0.29$\pm$0.07& 5.83 &  0.46$\pm$0.14& 0.60$\pm$0.11& \\
           &               &         &                    &              &      &               &              & \\
H$^{34}$SCN&$3_{03}-2_{02}$& 2-1     & 33642.089$\pm$0.020& 0.39$\pm$0.07& 5.83 &  0.91$\pm$0.17& 0.40$\pm$0.07& \\
           &$3_{03}-2_{02}$& Main    & 33642.328$\pm$0.010& 0.98$\pm$0.07& 5.83 &  0.96$\pm$0.09& 0.96$\pm$0.07& \\
           &$4_{04}-3_{03}$& Main    & 44856.085$\pm$0.020& 0.64$\pm$0.05& 5.83 &  0.57$\pm$0.06& 1.06$\pm$0.11& \\
           &               &         &                    &              &      &               &              & \\
NCS        & 5/2-3/2 ef & 9/2-7/2    & 42699.184$\pm$0.010& 2.25$\pm$0.07& 5.83 &  0.67$\pm$0.02& 3.14$\pm$0.10& \\
           & 5/2-3/2 ef & 7/2-4/2    & 42702.930$\pm$0.010& 2.10$\pm$0.08& 5.83 &  0.82$\pm$0.04& 2.39$\pm$0.10& \\
           & 5/2-3/2 ef & 5/2-3/2    & 43705.851$\pm$0.010& 1.61$\pm$0.08& 5.83 &  0.81$\pm$0.05& 1.88$\pm$0.10& \\
           &            &            &                    &              &      &               &              & \\
\hline
\end{tabular}
\tablefoot{
\tablefoottext{a}{Main refers to the three hyperfine components with the strongest line intensities, $F_u-F_l$=$J+1$$\rightarrow$$J$, 
$J$$\rightarrow$$J-1$, and $J-1$$\rightarrow$$J-2$, respectively. These are unresolved but could produce a measurable line broadening.
However, the model to compute the synthetic spectrum of all CHNS isomers and isotopologues takes into account all 
hyperfine components.}
\tablefoottext{b}{Observed frequencies for the transitions of the species
studied in this work assuming a v$_{LSR}$ of 5.83 \kms\, (see text). For lines for which only upper limits
to the intensity are obtained the predicted frequencies from the laboratory data are given.}
\tablefoottext{c}{Integrated line intensity in mK\,km\,s$^{-1}$.} 
\tablefoottext{d}{Line width at half intensity using a Gaussian fit in the line profile (in km~s$^{-1}$).}
\tablefoottext{e}{Antenna temperature (in mK). Upper limits correspond to 3$\sigma$ values.}
\tablefoottext{A}{These hyperfine components appear too strong compared with the other transition of the
same molecule. They are probably blended with unknown lines as shown in Fig. \ref{fig:thios}.}
\tablefoottext{B}{Heavily blended line. Derived parameters are uncertain (see Fig. \ref{fig:thios_hscn}).}
\tablefoottext{C}{The hyperfine components are blended but produce a well-defined line profile. The
line width has been fixed to 0.6 km\,s$^{-1}$ (see Fig. \ref{fig:thios_hscn}).}
\tablefoottext{D}{The main component is affected by a line of HC$_4$N at +8 MHz in the frequency switching
data with a throw of 8 MHz, and by a line of c-C$_3$HCCH at -10 MHz in the data with a throw of 10 MHz. 
The corresponding line parameters are uncertain (see Fig. \ref{fig:iso_hscn}).}
}
\end{table*}

\begin{table*}[h]
\centering
\caption{Observed line parameters for the isomers and isotopologues of HNCO}
\label{line_parameters_hnco}
\begin{tabular}{lcccrccrc}
\hline
Species    & $J_u-J_l$  &$F_u-F_l$$^a$& $\nu_{rest}$~$^b$ & $\int T_A^* dv$~$^c$ & v$_{LSR}$       & $\Delta v$~$^d$ & $T_A^*$~$^e$& Notes\\
                &                      & & (MHz)              & (mK\,km\,s$^{-1}$)  & (km\,s$^{-1}$)  & (km\,s$^{-1}$)  & (mK) & \\
\hline
HNCO       &$2_{02}-1_{01}$&1-1   & 43962.015$\pm$0.010 & 21.16$\pm$0.10& 5.83 &  0.63$\pm$0.01& 31.17$\pm$0.14& \\
           &               &1-2   & 43962.626$\pm$0.010 &  1.15$\pm$0.09& 5.83 &  0.58$\pm$0.05&  1.86$\pm$0.14& \\
           &               &3-2   & 43963.013$\pm$0.010 &174.41$\pm$0.01& 5.83 &  0.72$\pm$0.01&229.15$\pm$0.14&A\\
           &               &2-1   &                     &               &      &               &               &A\\
           &               &1-0   & 43963.556$\pm$0.010 & 29.93$\pm$0.12& 5.83 &  0.67$\pm$0.05& 42.19$\pm$0.14& \\
           &               &2-2   & 43963.659$\pm$0.010 & 18.87$\pm$0.11& 5.83 &  0.59$\pm$0.05& 29.71$\pm$0.14& \\
           &$4_{04}-3_{03}$&3-3   & 87924.350$\pm$0.010 & 10.42$\pm$1.68& 5.83 &  0.58$\pm$0.10& 16.89$\pm$2.02& \\
                   &               &Main  & 87925.238$\pm$0.010 &307.62$\pm$1.58& 5.83 &  0.63$\pm$0.01&525.21$\pm$2.02& \\         
           &               &4-4   & 87925.921$\pm$0.010 &  8.94$\pm$1.53& 5.83 &  0.60$\pm$0.11& 14.12$\pm$2.02& \\
           &$5_{05}-4_{04}$&Main  &109905.734$\pm$0.010 &175.38$\pm$1.31& 5.83 &  0.51$\pm$0.01&324.74$\pm$3.67& \\         
                   &               &      &                    &               &      &               &              & \\
HN$^{13}$CO&$2_{02}-1_{01}$&3-2   & 43964.265$\pm$0.010 &  3.41$\pm$0.04& 5.83 &  0.74$\pm$0.02&  4.33$\pm$0.14&A\\
           &               &2-1   &                     &               &      &               &               &A\\
           &               &1-0   & 43964.807$\pm$0.010 &  0.58$\pm$0.03& 5.83 &  0.57$\pm$0.04&  0.94$\pm$0.14& \\
           &               &2-2   & 43964.897$\pm$0.010 &  0.29$\pm$0.04& 5.83 &  0.48$\pm$0.15&  0.70$\pm$0.14& \\
           &               &      &                     &               &      &               &              & \\
H$^{15}$NCO&$2_{02}-1_{01}$&      & 42646.445$\pm$0.010 &  0.73$\pm$0.09& 5.83 &  0.59$\pm$0.09&  1.15$\pm$0.10& \\
           &            &         &                     &              &      &               &              & \\
DNCO       &$2_{02}-1_{01}$& 1-1  & 40785.534$\pm$0.010 &  0.54$\pm$0.10& 5.83 &  0.58$\pm$0.11&  0.87$\pm$0.07&\\ 
           &               & 3-2  & 40786.545$\pm$0.015 &  1.65$\pm$0.10& 5.83 &  0.60$\pm$0.15&  2.12$\pm$0.07&B\\
           &               & 2-1  & 40786.618$\pm$0.015 &  1.28$\pm$0.10& 5.83 &  0.60$\pm$0.15&  1.98$\pm$0.07&B\\
           &               & 1-0  & 40787.139$\pm$0.015 &  0.66$\pm$0.08& 5.83 &  0.60$\pm$0.15&  0.79$\pm$0.07&B\\
           &               & 2-2  & 40787.249$\pm$0.015 &  0.48$\pm$0.08& 5.83 &  0.60$\pm$0.15&  0.69$\pm$0.07&B\\
           &$4_{04}-3_{03}$& Main & 81571.895$\pm$0.010 &  6.34$\pm$0.37& 5.83 &  0.57$\pm$0.02& 10.42$\pm$0.66& \\
           &            &         &                     &              &      &               &              & \\
HCNO       & 2-1        & Main    & 45876.059$\pm$0.010 &  7.16$\pm$0.12& 5.83 &  0.91$\pm$0.02&  7.41$\pm$0.15&C\\         
           & 4-3        & Main    & 91751.350$\pm$0.010 &  2.11$\pm$0.60& 5.83 &  0.30$\pm$0.10&  6.60$\pm$1.68&C,D\\ 
           &            &         &                     &               &      &               &              & \\
DCNO       & 2-1        & Main    & 41169.798$\pm$0.010 &  0.25$\pm$0.05& 5.83 &  0.54$\pm$0.10&  0.44$\pm$0.08&C\\         
           &            &         &                     &               &      &               &              & \\
HOCN       &$2_{02}-1_{01}$& 2-2  & 41949.966$\pm$0.010 &  1.54$\pm$0.09& 5.83 &  0.61$\pm$0.05&  2.36$\pm$0.15& \\ 
           &               & 1-0  & 41950.108$\pm$0.010 &  2.14$\pm$0.09& 5.83 &  0.64$\pm$0.04&  3.14$\pm$0.15& \\ 
           &               & 2-1  & 41950.848$\pm$0.010 &  5.22$\pm$0.25& 5.83 &  0.70$\pm$0.15&  6.79$\pm$0.15&B\\ 
           &               & 3-2  & 41950.902$\pm$0.010 &  6.97$\pm$0.25& 5.83 &  0.70$\pm$0.15&  9.99$\pm$0.15&B\\ 
           &               & 1-1  & 41952.292$\pm$0.010 &  1.78$\pm$0.20& 5.83 &  0.69$\pm$0.06&  2.45$\pm$0.15& \\
           &$4_{04}-3_{03}$&Main  & 83900.560$\pm$0.010 & 12.77$\pm$0.53& 5.83 &  0.62$\pm$0.03& 19.35$\pm$0.61& \\
           &$5_{05}-4_{04}$&Main  &104874.696$\pm$0.010 &  4.44$\pm$0.77& 5.83 &  0.32$\pm$0.06& 13.11$\pm$2.53& \\
           &            &         &                     &               &      &               &              & \\
NCO        &7/2-5/2e    &9/2-7/2  &     81404.306$\pm$0.015  &  2.08$\pm$0.26& 5.83 &0.67$\pm$ 0.10 & 2.92$\pm$0.53& \\
           &7/2-5/2f    &9/2-7/2  &     81404.815$\pm$0.015  &  1.40$\pm$0.23& 5.83 &0.51$\pm$ 0.10 & 2.55$\pm$0.53& \\    
           &7/2-5/2e    &7/2-5/2  &     81413.143$\pm$0.020  &  0.81$\pm$0.25& 5.83 &0.47$\pm$ 0.14 & 1.62$\pm$0.65& \\    
           &7/2-5/2f    &7/2-5/2  &     81413.591$\pm$0.020  &  0.66$\pm$0.21& 5.83 &0.32$\pm$ 0.14 & 1.93$\pm$0.65& \\    
\hline
\end{tabular}
\tablefoot{
\tablefoottext{a}{Main refers to the three hyperfine components with the strongest line intensities, $F_u-F_l$=$J+1$$\rightarrow$$J$, 
$J$$\rightarrow$$J-1$, and $J-1$$\rightarrow$$J-2$, respectively. These are unresolved but could produce a measurable line 
broadening. However, the model
to compute the synthetic spectrum of all CHNO isomers and isotopologues takes into account all 
hyperfine components.}
\tablefoottext{b}{Predicted, or observed, frequencies for the transitions of the species
studied in this work (see text).}
\tablefoottext{c}{Integrated line intensity in mK\,km\,s$^{-1}$.} 
\tablefoottext{d}{Line width at half intensity using a Gaussian fit in the line profile (in km~s$^{-1}$).}
\tablefoottext{e}{Antenna temperature (in mK).}
\tablefoottext{A}{The $F=3-2$ and $F=2-1$ hyperfine componentes are blended with a predicted
separation between them of 40 kHz. The derived frequency corresponds to that of the averaged value of both components.}
\tablefoottext{B}{The hyperfine components are blended. The line width has been fixed. We estimate
an uncertainty on this parameter of 0.15 \kms.}
\tablefoottext{C}{The hyperfine components are blended. A single line has been fitted (see Fig. \ref{fig:HNCO}).}
\tablefoottext{D}{Blended with a U-line as shown in Fig. \ref{fig:thios_hscn} but line parameters
can be still estimated.}
}
\end{table*}

\end{appendix}
\end{document}